\documentclass[runningheads]{llncs}
\usepackage{algorithm}
\usepackage{algorithmic}
\usepackage{makecell}
\usepackage{pifont}
\usepackage{multirow}
\usepackage{amsmath}
\usepackage[hyphens]{url}
\usepackage{placeins}
\usepackage{amsfonts}
\usepackage{booktabs}
\usepackage{graphicx}

\begin{document}
\title{Joint Latent Topic Discovery and Expectation Modeling for Financial Markets}

\author{Lili Wang\inst{1} \and
Chenghan Huang\inst{2} \and
Chongyang Gao\inst{3} \and
Weicheng Ma \inst{1} \and
Soroush Vosoughi\inst{1}}
\authorrunning{L. Wang et al.}

\institute{Dartmouth College, Hanover NH 03755, USA \and
Millennium Management, LLC, New York NY 10022, USA \and
Northwestern University, Evanston IL 60208, USA 
\email{\{lili.wang.gr,soroush\}@dartmouth.edu}}

\titlerunning{Joint Latent Topic Discovery and Expectation Modeling}

\maketitle            
\begin{abstract}
In the pursuit of accurate and scalable quantitative methods for financial market analysis, the focus has shifted from individual stock models to those capturing interrelations between companies and their stocks. However, current relational stock methods are limited by their reliance on predefined stock relationships and the exclusive consideration of immediate effects. To address these limitations, we present a groundbreaking framework for financial market analysis. This approach, to our knowledge, is the first to jointly model investor expectations and automatically mine latent stock relationships. Comprehensive experiments conducted on China's CSI 300, one of the world's largest markets, demonstrate that our model consistently achieves an annual return exceeding 10\%. This performance surpasses existing benchmarks, setting a new state-of-the-art standard in stock return prediction and multiyear trading simulations (i.e., backtesting).
\keywords{stock trend prediction  \and  trading simulation \and expectation modeling.}
\end{abstract}

\section{Introduction}
The efficient-market hypothesis in traditional finance posits that stock prices reflect all available market information, with current prices consistently trading at their fair value \cite{fama1965behavior}. Consequently, predicting future stock prices is challenging without access to new information. However, markets are often less efficient in reality \cite{poterba1988mean}, with stock market fluctuations driven by behavioral factors such as expectations, confidence, panic, euphoria, or herding behavior. These inefficiencies enable the use of machine learning to predict future stock movements based on historical trends.

Stock-affecting behavioral factors can be categorized into short- and long-term factors. Factors like panic, euphoria, or herding behavior are typically short-term, while subjective expectations and confidence tend to be long-term factors, only influencing stock prices imperceptibly over extended periods. These factors do not solely impact individual stocks; their effects often spread to topically related stocks, which share similarities across various explicit or latent dimensions. Recent stock prediction works \cite{sawhney2021stock,HIST,sawhney2020deep} utilize topic stocks to improve prediction capabilities. However, most of these methods exhibit two key limitations:

(1) Topics are typically assumed to be static and known beforehand. However, real-world topics can change and new topics may emerge. For example, during the COVID-19 pandemic, pharmaceutical companies investing in COVID vaccines (e.g., Pfizer\footnote{\url{https://investors.pfizer.com/Investors/Stock-Info/default.aspx}} and Moderna\footnote{\url{https://investors.modernatx.com/Stock-Info/default.aspx}}) experienced stock price fluctuations under the new COVID topic.

\begin{figure}[t]
\centering
\includegraphics[width=0.80\columnwidth]{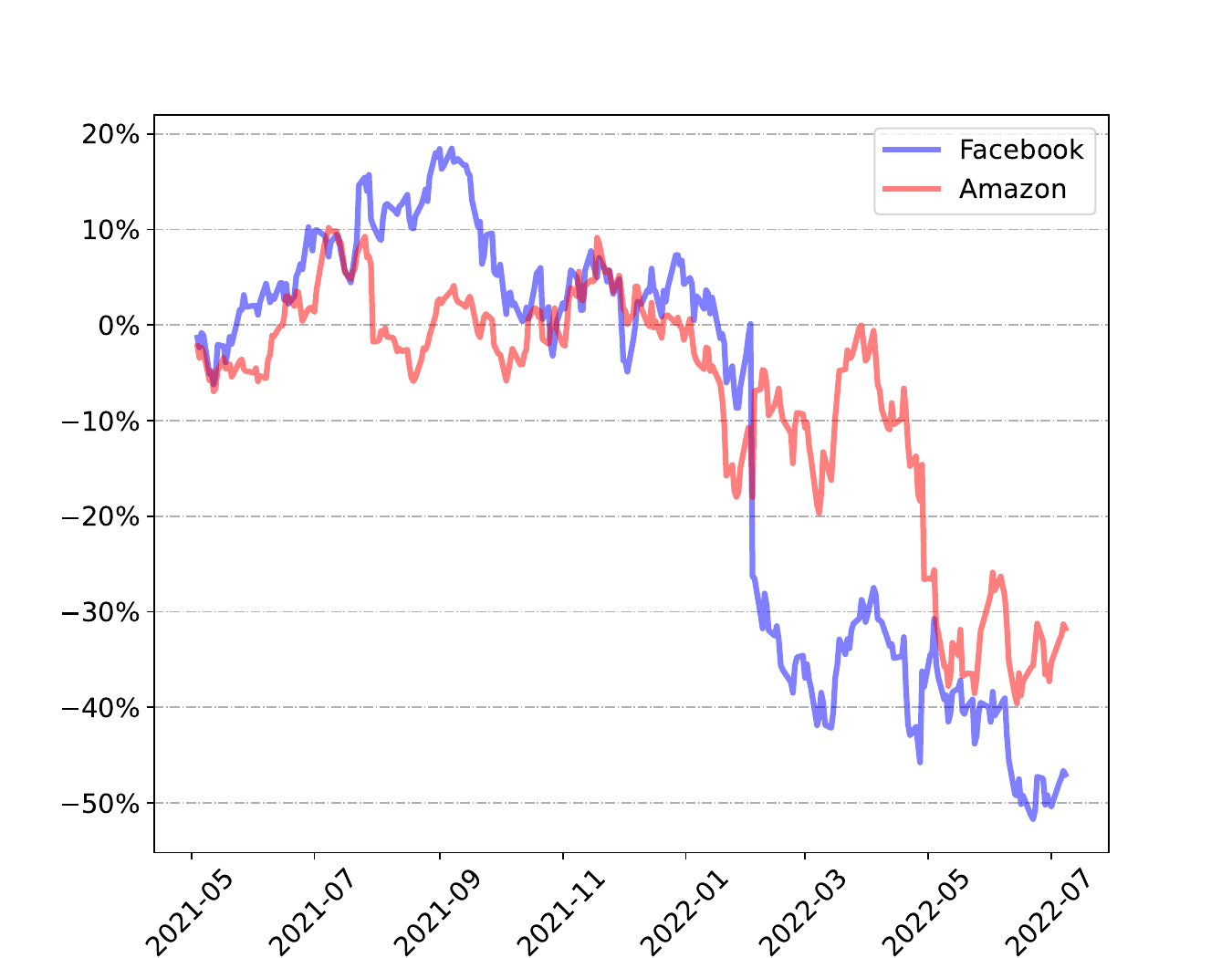} 
\caption{The return of Amazon and Facebook (Meta) stocks from 2021-05-03 to 2022-07-08 with respect to their stock prices at 2021-05-03.}.
\label{FVA}
\end{figure}

(2) Only the \emph{short-term impact} between stocks is considered, neglecting the \emph{long-term subjective expectations}. Unlike analyst expectations, subjective expectations are based on human psychology and behavior and can be irrational. Figure \ref{FVA} illustrates that Amazon and Facebook stock prices often correlate, and previous methods might reason that a significant drop in Facebook's price would also lead to plummeting Amazon stocks. However, in the second half of 2021, Amazon's return was lower than Facebook's, lowering investor expectations for Amazon. Thus, when Amazon released an unremarkable financial report\footnote{\url{https://s2.q4cdn.com/299287126/files/doc_financials/2021/q4/business_and_financial_update.pdf}} on February 3, 2022, its stock rose 13.5

In this paper, we introduce a novel attention-based framework for stock trend prediction that simultaneously discovers topical relations between stocks and models both the \emph{short-term impact} and \emph{long-term subjective expectations} of topically similar stocks. To the best of our knowledge, our framework is the first to:

\begin{itemize}
\item Model the influence of investors' subjective expectations on stock prices.

\item Automatically identify dynamic topics between stocks without making assumptions or requiring additional knowledge.
\end{itemize}

Through comprehensive experiments against 16 well-established baselines, we demonstrate that our method achieves the current state-of-the-art on the Qlib \cite{qlib} quantitative investment platform.

\section{Related Work}
The stock price prediction and stock selection problems can be easily formed as a time series forecasting problem. Therefore,  traditional and deep-learning-based machine learning (ML) methods, especially those for sequence learning, have been directly applied to these tasks are widely used by investment institutions. Specifically, Qlib \cite{qlib}, a popular quantitative investment platform, benchmarks models based on the following ML methods:  multi-layer perceptron (MLP); TabNet \cite{tabnet}; TCN \cite{TCN}; gradient boosting models: CatBoost \cite{CatBoost}, LightGBM \cite{LightGBM}; Recurrent Neural Network (RNN) based models: long short-term memory (LSTM) \cite{LSTM}, gated recurrent unit (GRU) \cite{GRU}, DA-RNN \cite{DARNN}, AdaRNN \cite{adarnn}; and attention-based models: Transformer \cite{Transformer}, and Localformer \cite{localformer}. To model the co-movement and relations among stocks, some research, such as MAN-SF \cite{sawhney2020deep} and STHAN-SR \cite{sawhney2021stock}), also adopted graph neural network methods like GCN \cite{gcn} and GATs \cite{gat} to mine the correlation between different stocks.

More recent models include those specifically designed for stock trading. DoubleEnsemble \cite{Doubleensemble} is an ensemble model which utilizes learning-trajectory-based sample reweighting and shuffling-based feature selection for stock prediction. ADD \cite{ADD} attempt to extract clean information from noisy data to improve prediction performances. Specifically, they proposed a method for separating the inferential features from the noisy raw data to a certain degree using disentanglement, dynamic self-distillation, and data augmentation. Xu et al. assume that inter-dependencies may exist among different stocks at different time series and propose a method called IGMTF \cite{IGMTF} to mine these relations. In their other work, they propose HIST \cite{HIST}, a three-step framework to mine the concept-oriented shared information and individual features among stocks.

We use most of the above-mentioned methods as baselines in our experiments.

\section{Framework}

\subsection{Problem Definition}

We formulate the stock trend prediction problem as a regression problem. Let $stock_1$, $stock_2$, ..., $stock_n$ denote $n$ different stocks. For each stock $stock_j$ on date $i$, the closing price is $price_j^i$. Given the historical information before date $i$, our task is to predict the one-day return $r^{i}_{j}=\frac{price^{i}_{j}-price^{i-1}_{j}}{price^{i-1}_{j}}$ for each stock $j$ on date $i$. In the rest of this paper, we use $r^i$ to denote ($r^i_1$, $r^i_2$, ..., $r^i_n$).
\begin{figure*}[htbp]
\centering
\includegraphics[width=0.90\columnwidth]{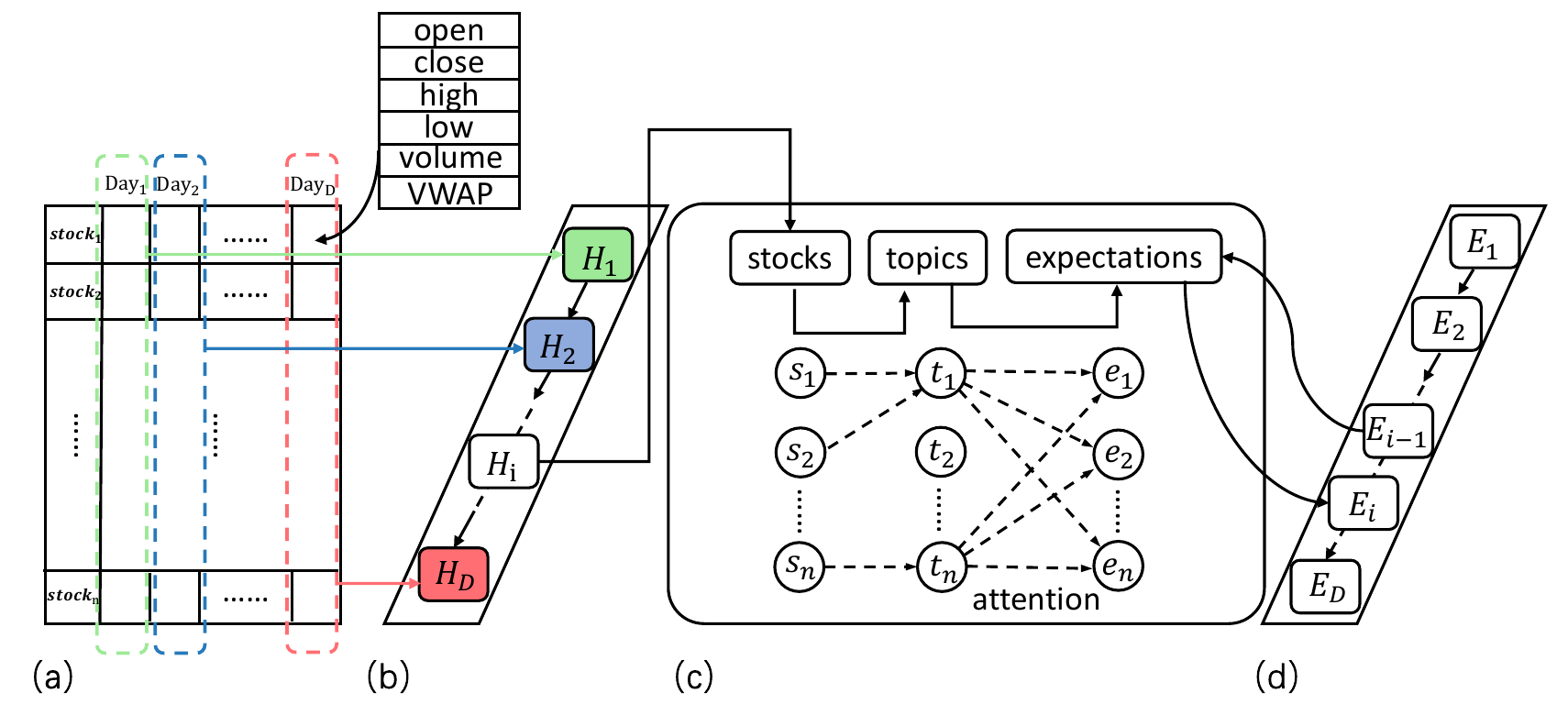} 
\caption{Our model's framework consists of: (a) Extracting Alpha360 features from raw data: For each stock on a given day, we combine the opening price, closing price, highest price, lowest price, trading volume, and volume-weighted average price (VWAP) into a 6-D feature vector. We then concatenate this vector with similar 6-D vectors from the preceding 59 days to form a 360-D feature vector. (b) The LSTM module processes the extracted Alpha360 features to learn temporal representations. (c) The left half of this section represents the topic module, which uses stock embeddings as input to extract latent topics. The right half illustrates the expectations module, which takes the $E_{i-1}$ output from the expectation LSTM in part (d) as an initial embedding, employs attention with topics to update it to $\hat E_{i-1}$, and feeds it back into the expectation LSTM as input for day $i$. (d) The second LSTM module models the evolution of each stock's expectation.}
\label{framework}
\end{figure*}
\subsection{Overview of the Framework}
The architecture of our model is shown in Figure \ref{framework}. The model consists of three jointly optimized modules: temporal stock representation (which aims to extract temporal stock features), topic module (aims to discover the dynamic topics based on the extracted features), and expectation module (aims to model the subjective expectations for each stock). Below we describe each module in detail.

\subsection{Temporal Stock Representation}
The first step of our learning framework is to extract the Alpha360 features \cite{qlib} from the raw data. The Alpha360 is a 360-D feature vector that is widely used in the quantitative investment domain. As shown in Figure \ref{framework} (a), for each stock on each day, we combine the opening price, closing price, highest price, lowest price, trading volume and volume-weighted average price (VWAP) as a 6-D feature vector and concatenate it with similar 6-D vectors from the past 59 days to get a 360-D feature vector. 

To extract the temporal representation of stocks, we adopt an LSTM layer shown in Figure \ref{framework} (b)). Our framework is trained recursively by date: for each trading day $i$, the input is the Alpha360 features $H_i$ of $stock_1$, $stock_2$, ..., $stock_n$ for that day and the output of the LSTM layer is $S_i$, which is comprised of $s^i_1$, $s^i_2$, ..., $s^i_n$, denoting the embeddings of each stock. 

\subsection{Topic Module}
As mentioned before, the relations among stocks may evolve overtime, so our framework needs to be able to capture the evolution of topics and discover new topics each day. Figure \ref{framework} (c) shows the topic and expectation modules of our framework.

First, for each day $i$, we initialize the $n$ topic embeddings $T_i=(t^i_1$, $t^i_2$, ..., $t^i_n)$ using the $n$ stock embeddings $S_i=(s^i_1$, $s^i_2$, ..., $s^i_n)$. Then, we compute the Tanimoto coefficient ($\mathcal{T}$) \cite{tanimoto1958elementary} between all pairs of $t^i_{j_1}$ (topic $j_1$ in day $i$) and $s^i_{j_2}$ (stock $j_2$ in day $i$), for $\forall j_1,j_2 \in [1,n] $ with the following equation:

\begin{equation}
\small
\mathcal{T}(t^i_{j_1}, s^i_{j_2})=\frac{t^i_{j_1} s^i_{j_2}}{\|t^i_{j_1}\|^{2}+\|s^i_{j_2}\|^{2}-t^i_{j_1} s^i_{j_2}}
\label{Tanimoto}
\end{equation}

We define a function $\phi^{i}(s^i_{j_2})$ that for each stock embedding, $s^i_{j_2}$, returns the most similar topic index $j_1$, except for its own topic (i.e., $j_1 \neq j_2$) in date $i$, based on the Tanimoto coefficient:
\begin{equation}
\small
\phi^{i}(s^i_{j_2})=\arg \max _{j_1} \left(\mathcal{T}(t^i_{j_1}, s^i_{j_2}), j_1 \neq j_2 \right)
\end{equation}
In the example shown in Figure \ref{framework} (c) (with the dashed lines) $\phi^{i}(s^i_{1})=1,$ $\phi^{i}(s^i_{2})=1,$ $\phi^{i}(s^i_{n})=n.$ 

We further construct a set $valid^i$ that contains ``valid'' topics for each day $i$, i.e., those that are the most related to at least one stock:
\begin{equation}
\small
    valid^i= \left\{x | \exists j, x = \phi^{i}(s^i_{j}) \right\}
    \label{valid}
\end{equation}

This set denotes the topics we discovered for each day. Only if a topic $t^i_{j_1}$ is the most similar topic to at least one stock, it will be include in this set, other topics (e.g., $t^i_2$ in Figure \ref{framework} (c)) will be excluded from the following calculations.

To update each topic embedding $t^i_{j_1} (j_1 \in valid^i)$, we train the fully connected layer with weight matrix $W_t$, bias matrix $b_t$ and activation function $tanh$ to aggregate the stock embeddings using the Tanimoto coefficient:

\begin{equation}
\small
t^i_{j_1}=tanh\left(W_{t}\left(\sum_{\phi^{i}(s^i_{j_2})=j_1} \mathcal{T}(t^i_{j_1}, s^i_{j_2})s^i_{j_2}  \right)+b_{t}\right)
\label{topic_update}
\end{equation}

\subsection{Expectation Module}
The expectations of investors change over time and our framework needs to take that into consideration.  As shown in Figure \ref{framework} (d), we adopt an LSTM to model the evolving expectations of each stock. Each $E_i$ consists of $n$ expectation embeddings $e^i_1$, $e^i_2$, ..., $e^i_n$, we assume that at the first timestamp, the investor's expectations are all decided by the stocks themselves, so the initial embedding $E_1=(e^1_1, e^1_2, ..., e^1_n)$ are initialized as the $n$ stock embeddings $S_1=(s^1_1, s^1_2, ..., s^1_n)$.

The expectation for one stock can also be affected by the performance of other stocks under related topics. So for day $i$, we take the output $E_{i-1}=(e^{i-1}_1, e^{i-1}_2, ..., e^{i-1}_n)$ of the LSTM and   
adopt an attention mechanism to learn the importance of each topic $j_1$ to the expectations:

\begin{equation}
\small
\alpha(t^i_{j_1}, e^{i-1}_{j_2})=\frac{\exp \left(   \mathcal{T}\left( t^i_{j_1}, e^{i-1}_{j_2}\right)   \right)}{\sum_{j \in valid^i} \exp \left(   \mathcal{T}\left( t^i_{j}, e^{i-1}_{j_2}\right)   \right)}
\label{alpha}
\end{equation}

\begin{equation}
\small
\hat e^{i-1}_{j_2}=tanh\left(W^1_{e}e^{i-1}_{j_2}+ W^2_{e}\left(\sum_{j_1 \in valid^i} \alpha(t^i_{j_1}, e^{i-1}_{j_2})t^i_{j_1}  \right)+b_{e}\right)
\label{attention_e}
\end{equation}

where $\alpha(t^i_{j_1}, e^{i-1}_{j_2})$ measures the importance of topic $j_1$ to the expectation of stock $j_2$, and the updated $\hat E_{i-1}=(\hat e^{i-1}_1, \hat e^{i-1}_2, ..., \hat e^{i-1}_n)$ then feed back to the LSTM (d) as the input of day $i$.

\subsection{Loss Function}
The objective of our model is to predict the one-day return $r$ of each stock. The objective relies on three components: $r_{stock}$, $r_{topic}$, and $r_{expectation}$. 

The $r_{stock}$ and $r_{expectation}$ are learnt from the temporal stock embeddings and the expectation embeddings, respectively :
\begin{equation}
\small
r^{i}_{stock}=tanh\left(W_{stock} S_{i}+b_{stock}\right)
\end{equation}

\begin{equation}
\small
r^{i}_{expectation}=tanh\left(W_{expectation} E_{i}+b_{expectation}\right)
\end{equation}

To learn $r_{topic}$, we first learn the importance of each topic to the stocks using a similar attention mechanism as the expectation module:

\begin{equation}
\small
\beta(t^i_{j_1}, s^{i}_{j_2})=\frac{\exp \left(   \mathcal{T}\left( t^i_{j_1},  s^{i}_{j_2}\right)   \right)}{\sum_{j \in valid^i} \exp \left(   \mathcal{T}\left( t^i_{j},  s^{i}_{j_2}\right)   \right)}
\end{equation}

\begin{equation}
\small
o^{i}_{j_2}=tanh\left( W_{s}\left(\sum_{j_1 \in valid^i} \beta(t^i_{j_1},  s^{i}_{j_2})t^i_{j_1}  \right)+b_{s}\right)
\end{equation}

where $\beta(t^i_{j_1}, s^{i}_{j_2})$ measures the importance of topic $j_1$ to the expectation of stock $j_2$ on day $i$. Note that different from the expectation module which includes the term $W^1_{e}e^{i-1}_{j_2}$, here $o^{i}_{j_2}$ measures the impact of all the topics on the stock $j_2$ on day $i$, without considering $s_i$. This is because $s_i$ is already included in $r_{stock}$. We use $O^i$ to denote $(o^i_1$, $o^i_2$, ..., $o^i_n)$; $r_{topic}$ is learnt as:

\begin{equation}
\small
r^{i}_{topic}=tanh\left(W_{topic} O^{i}+b_{topic}\right)
\end{equation}

The predicted return $\hat r$ is learnt by combining these three components:

\begin{equation}
\small
\hat r^{i}=tanh\left(W_{\hat r}  r_{stock}^{i}+W_{\hat r} r_{topic}^{i}+W_{\hat r}r_{expectation}^{i}+b_{\hat r}\right)
\label{pred_r}
\end{equation}

The loss function of our model is defined as the 
mean squared error between $\hat r$ and $r$:

\begin{equation}
\small
\mathcal{L}=\frac{\sum_{i \in [1,D]}  \left(r^{i}-\hat r^{i}\right)^{\top} \left(r^{i}-\hat r^{i}\right)}{D \cdot n}
\label{loss}
\end{equation}
where D corresponds to the number of trading days. Algorithm \ref{alg:algorithm} shows the
pseudocode of our method.

\begin{algorithm}[ht]
\small
\caption{Training pseudo-code}
\label{alg:algorithm}
\textbf{Input}: $H =\left\{H_{1}, H_{2}, \ldots, H_{|D|}\right\}$: the Alpha360 features for each trading day\\

\textbf{Parameters}: \\
 $\Theta:$ the initialized model parameters, $epochs :$ the number of training epochs, $\eta$ : learning rate\\
 
\textbf{Output}: The predicted return $\hat r$ \\

\begin{algorithmic}[1] 
\FOR{$epoch \leftarrow \{1, \ldots, epochs\}$} 
\FOR{$i \leftarrow \{1, \ldots, D\}$} 
\STATE $S_i \leftarrow  LSTM_b (H_i)$
\IF {$t == 1$}
\STATE $T_i \leftarrow S_i$
\STATE $E_i \leftarrow S_i$
\ENDIF
\STATE $\mathcal{T} \leftarrow$ Calculate Tanimoto coefficient (Eq. 1)

\STATE $valid^i \leftarrow$ Calculate the valid topic set according to $\mathcal{T}$ (Eq. 3)

\STATE $T_i \leftarrow$   Aggregate information from $S_i$ according to $\mathcal{T}$ (Eq. 4)

\STATE $\alpha \leftarrow$ Calculate the attention weight  (Eq. 5)

\STATE $\hat E_{i} \leftarrow$ Aggregate information from $T_i$ according to $\alpha$ (Eq. 6)

\STATE $E_{i+1} \leftarrow  LSTM_d (\hat E_{i})$
\ENDFOR

\STATE Compute the stochastic gradients of $\Theta$ (Eq.13)

\STATE Update model parameters $\Theta$ according to learning rate $\eta$ and gradients.

\ENDFOR

\STATE \textbf{return} the predicted return $\hat r$
\end{algorithmic}
\end{algorithm}

\subsection{Model Training}\label{training}
Our model is optimized by minimizing the global loss $\mathcal{L}$. This was done using the Adam optimizer \cite{kingma2014adam}. The hyper-parameters are set as follows: the embedding size is set to 128, the learning rate is set to 0.001, the training epoch is set to 300, the dropout rate is set to 0.1. All experiments are run on a Lambda Deep Learning 2-GPU Workstation (RTX 2080) with 24GB of memory, and the random seed is set to 0 at the beginning of each experiment.


\section{Experiments}
\subsection{Datasets}
We run comprehensive evaluations of our framework on the China's CSI 300 financial markets, from 2008 to 2022. We use the data from 01/01/2008 to 12/31/2014 as the training set, the data from 01/01/2015 to 12/31/2016 as the validation set for hyper-parameter fine-tuning, and the data from 01/01/2017 to 07/10/2022 as the test set.

\subsection{Baselines}
We compare our framework with a comprehensive list of 16 well-known methods which are widely used in the financial sector. These methods span six different categories and are:
\begin{itemize}

\item \noindent\textbf{Classic Models -}
MLP,  TCN \cite{TCN},   GATs \cite{gat}

\item \noindent\textbf{Tabular Learning -} TabNet

\item \noindent\textbf{Gradient Boosting Models -}  CatBoost \cite{CatBoost}, LightGBM \cite{LightGBM}

\item \noindent\textbf{RNN-Based Methods-} LSTM \cite{LSTM}, GRU \cite{GRU}, DA-RNN \cite{DARNN}, AdaRNN \cite{adarnn}

\item \noindent\textbf{Attention-Based Methods-} Transformer \cite{Transformer}, Localformer \cite{localformer}

\item \noindent\textbf{Financial Prediction Methods-} DoubleEnsemble \cite{Doubleensemble}, ADD \cite{ADD}, HIST \cite{HIST}, IGMTF \cite{IGMTF} 

\end{itemize}

Note that although our method can mine the latent topics among stocks, the tasks in our experiments only assume access to price and volume features (opening price, closing price, highest price, lowest price, VWAP). Several recently proposed methods require additional information such as company relations \cite{sawhney2021stock} or social media text \cite{sawhney2020deep}, thus these methods cannot be included as baselines.

\begin{table*}[]
\centering
\begin{tabular}{ccccc}
\Xhline{2\arrayrulewidth}
\multicolumn{1}{c}{\textbf{Model Name}} & \multicolumn{1}{c}{\textbf{IC}} & \multicolumn{1}{c}{\textbf{ICIR}} & \multicolumn{1}{c}{\textbf{Rank IC}} & \multicolumn{1}{c}{\textbf{Rank ICIR}} \\ \Xhline{2\arrayrulewidth}
Transformer       & 0.0143±0.0024     *                &   0.0910±0.0180         *              & 0.0317±0.0024   *                             & 0.2192±0.0190          *              \\

TabNet           &  0.0286±0.0000      *              & 0.1975±0.0000         *                &  0.0367±0.0000       *                          &  0.2798±0.0000       *              \\

MLP                                       & 0.0267±0.0017    *                & 0.1845±0.0154        *               &  0.0362±0.0018          *                       & 0.2681±0.0157        *     \\

Localformer         &  0.0358±0.0036        *             & 0.2633±0.0334         *                &  0.0477±0.0019   *                               &   0.3643±0.0218      *                     \\ 

CatBoost & 0.0326±0.0000       *               &  0.2328±0.0000         *                &  0.0394±0.0000      *                           & 0.2998±0.0000      *                   \\ 

DoubleEnsemble     & 0.0362±0.0005         *             & 0.2725±0.0036        *                & 0.0444±0.0004       *                          & 0.3450±0.0038         *                  \\ 

LightGBM               & 0.0347±0.0000     *                  &  0.2648±0.0000       *                &  0.0443±0.0000      *                      &0.3520±0.0000          *                  \\ 

TCN                 &  0.0384±0.0015      *                &  0.2834±0.0164     *                    &0.0455±0.0012    *                              & 0.3546±0.0077   *            \\ 

ALSTM                  &   0.0413±0.0034    *                &  0.3166±0.0329       *                &   0.0504±0.0032   *                             & 0.3974±0.0280     *      \\ 

LSTM            & 0.0402±0.0030      *              & 0.3194±0.0271          *            & 0.0496±0.0027           *                        &  0.4040±0.0212          *                    \\ 

ADD                                       &  0.0370±0.0025           *          & 0.2669±0.0254        *                & 0.0511±0.0018           *                      & 0.3756±0.0235       *     \\ 

GRU                & 0.0417±0.0029       *               &   0.3284±0.0367      *                   &  0.0510±0.0014   *                              &  0.4137±0.0224     *      \\ 

AdaRNN                &  0.0380±0.0117      *              & 0.2999±0.1022      *                  & 0.0472±0.0095      *                          &  0.3744±0.0974        *          \\

GATs         & 0.0430±0.0010     *                &   0.3221±0.0096        *               & 0.0543±0.0012      *                          &  0.4217±0.0099    *      \\

IGMTF                &  0.0419±0.0004    *             &  0.3152±0.0055      *                  & 0.0538±0.0014    *                               &  0.4213±0.0171       *            \\ 

HIST               & 0.0437±0.0012     *                 & 0.2952±0.0108       *                 & 0.0581±0.0013     *                            &  0.3912±0.0096    *       \\ 

Our Method &  \textbf{0.0489±0.0026}     &  \textbf{0.3593±0.0143}    & \textbf{0.0605±0.0023}  & \textbf{0.4514±0.0225}
\\ \Xhline{2\arrayrulewidth}
\end{tabular}

\caption{The results of stock trend prediction on the CSI300 market from 01/01/2017 to 07/10/2022. All the results are averaged after 10 runs, and the standard deviations are shown. * corresponds to statistically significant differences between a baseline and our method ($p < 0.05$ using \textup{t}-test).}
\label{trend}
\end{table*}

\begin{table*}[ht]
\centering
\begin{tabular}{cccc}
\Xhline{2\arrayrulewidth}
\multicolumn{1}{c}{\textbf{Model Name}} &  \multicolumn{1}{c}{\textbf{Annualized Return}} & \multicolumn{1}{c}{\textbf{Max Drawdown}} & \multicolumn{1}{c}{\textbf{Information Ratio}} \\ \Xhline{2\arrayrulewidth}
Transformer                            & 0.0069±0.0181   *                              & -0.2131±0.0868 *      & 0.0753±0.2138   *                    \\ 
TabNet                          &  0.0719±0.0000   *                               & -0.1139±0.0000     & 0.8155±0.0000    *               \\ 
MLP                                                     &  0.0441±0.0153  *                                & -0.1512±0.0375  *     &   0.5163±0.1882     *               \\
Localformer                        & 0.0498±0.0228   *                                &   -0.1268±0.0235    &  0.6194±0.2843   *                        \\ 
CatBoost                     &  0.0585±0.0013   *                              & -0.1364±0.0051     &    0.7270±0.0162   *                    \\ 
DoubleEnsemble                        & 0.0642±0.0112    *                                 & -0.0900±0.0103 *  & 0.8234±0.1398     *                      \\ 
LightGBM                            &  0.0707±0.0000   *                                  & \textbf{-0.0835±0.0000} * & 0.9487±0.0000  *                          \\ 
TCN                                     &0.0781±0.0203    *                               & -0.0849±0.0151 *          & 	 1.0205±0.2350  *             \\ 
ALSTM                                    &  0.0777±0.0220    *                                 & -0.1031±0.0204                   & 	 1.0226±0.2859    *       \\ 
LSTM                            & 0.0826±0.0242       *                            & -0.0908±0.0132 * &  1.0706±0.2771  *                            \\ 
ADD                                     & 0.0759±0.0178    *                               & -0.0939±0.0237           & 0.9471±0.2101 *                \\ 
GRU                             & 0.0815±0.0258    *                              &  -0.0917±0.0270    *             & 	1.0826±0.3671           \\ 
AdaRNN                         & 0.0619±0.0589      *                              &  -0.1392±0.1622                 & 0.8439±0.7172       \\ 
GATs                              & 0.0886±0.0115    *                                &  -0.1022±0.0184                   & 1.1524±0.1469   *       \\ 

IGMTF                              & 0.0903±0.0095     *                              &  -0.0986±0.0174                   & 1.1825±0.1035           \\ 
HIST                                   & 0.0854±0.0119     *                              &  -0.0919±0.0152  *                & 	 1.0879±0.1504 *         \\ 
Our Method  &  \textbf{0.1063±0.0187}                                   &   -0.1191±0.0301                 & 	  \textbf{1.3315±0.2169}  
\\ \Xhline{2\arrayrulewidth}
\end{tabular}
\caption{The results of trading simulation on the CSI300 market from 01/01/2017 to 07/10/2022. All the results are averaged after 10 runs, and the standard deviations are shown. * corresponds to statistically significant differences between a baseline and our method ($p < 0.05$ using \textup{t}-test).}
\label{simulation}
\end{table*}

\subsection{Results}
We use \textbf{stock trend prediction} and \textbf{trading simulation} for our experiments.

\subsubsection{Stock Trend Prediction} 
This task aims to evaluate the ability of models to predict the future stock price trend. For each trading day $i$, we calculate the 1-day return $\hat{r}^{i}$ of each stock based on its historical information before date $i$. For the results, we report the averaged information coefficient (IC), ranked information coefficient (Rank IC), information ratio of IC (ICIR), and information ratio of Rank IC (Rank ICIR). $\mathrm{IC}^{i}$ is the daily IC that measures the Pearson correlation between the predicted ratio $\hat{r}^{i}$ and the ground-truth ratio ${r}^{i}$:

\begin{equation}
\small
\mathrm{IC^{i}}= \frac{(\hat{r}^{i}-\operatorname{mean}(\hat{r}^{i}))^{\top}(r^{i}-\operatorname{mean}(r^{i}))}{n \cdot \operatorname{std}(\hat{r^{i}}) \cdot \operatorname{std}(r^{i})}
\end{equation}

The IC is calculated for the average of each trading day:

\begin{equation}
\small
\mathrm{IC}=\frac{ \sum_{i \in [1,D]} \mathrm{IC^{i}}}{D}
\end{equation}

The ICIR is used to show the stability of IC, which is calculated by dividing IC by its standard deviation:

\begin{equation}
\small
\mathrm{ICIR}= \frac{\mathrm{IC}}{\operatorname{std}(\mathrm{IC})}
\end{equation}

For the caculation of $\mathrm{Rank\ IC^{i}}$, we first use $R^{i}=rank(r^{i})$, and $\hat{R}^{i}=rank(\hat r^{i})$ to denote the ranks of the ground-truth and the predicted ratios, respectively:
\begin{equation}
\small
\mathrm{Rank\ IC^{i}}= \frac{(\hat{R}^{i}-\operatorname{mean}(\hat{R}^{i}))^{\top}(R^{i}-\operatorname{mean}(R^{i}))}{n \cdot \operatorname{std}(\hat{R^{i}}) \cdot \operatorname{std}(R^{i})}
\end{equation}

The Rank IC and Rank ICIR are calculated similarly as before:
\begin{equation}
\small
\mathrm{Rank\ IC}=\frac{ \sum_{i \in [1,D]} \mathrm{Rank\ IC^{i}}}{D}
\end{equation}

\begin{equation}
\small
\mathrm{Rank\ ICIR}= \frac{\mathrm{Rank\ IC}}{\operatorname{std}(\mathrm{Rank\ IC})}
\end{equation}

The results of the stock trend prediction task on the test set of the China CSI300 market (01/01/2017 to 07/10/2022) are shown in Table \ref{trend}. Our method significantly outperforms all the 16 baselines across all four metrics (IC, ICIR, Rank IC, and Rank ICIR) with around 10\% enhancement over the second-place model for each metric. These results indicate the importance of modeling expectations and dynamic topics in financial market analysis. It is also interesting to note that the traditional RNN-based methods (such as GRU and LSTM) achieve similar or even better results compared to the models specifically designed for financial analysis (such as ADD, IGMTF, and DoubleEnsemble). This may be attributed to the low signal-to-noise ratio in the financial market since the simpler models may be more robust to noise. These observations further demonstrate the hardness of this task.

\subsubsection{Trading Simulation}
In quantitative investment, "backtesting" refers to applying a trading strategy to historical data, simulating trading, and measuring the return of the strategy. For this task, we employ the top-$k$ dropout strategy for each method, reporting the \textbf{annualized return}\footnote{\url{https://www.investopedia.com/terms/a/annualized-rate.asp}} (the geometric average of money earned by an investment strategy each year over a given time period), \textbf{max drawdown}\footnote{\url{https://www.investopedia.com/terms/m/maximum-drawdown-mdd.asp}} (maximum observed loss from a peak to a trough), and the \textbf{information ratio}\footnote{\url{https://www.investopedia.com/terms/i/informationratio.asp}} (ratio of returns above the returns of the CSI300 benchmark). The top-$k$ dropout strategy is a straightforward quantitative investment approach: for each trading day, we hold $k$ stocks, sell $d$ stocks with the worst predicted 1-day return, and buy $d$ unheld stocks with the best-predicted 1-day return. In our experiments, $k$ is set to 50, and $d$ is set to 5. The trading simulation task results on the test set of the China CSI300 market are displayed in Table \ref{simulation}. Our method surpasses all 16 baselines in annualized return and information ratio. To improve the stability of profitability, future research could explore modifications designed to reduce the max drawdown of our approach.

\section{Conclusion}
In this paper, we introduce a novel framework for stock trend prediction, suitable for quantitative analysis of financial markets and stock selection. To the best of our knowledge, our method is the first to consider (1) investors' subjective expectations, and (2) automatically mined dynamic topics that do not require additional knowledge. Through experiments on 16 baselines using the CSI 300 market, we demonstrate that our model achieves a stable annual return above 10\%, outperforming all existing baselines and attaining the current state-of-the-art results for stock trend prediction and trading simulation tasks.

Future work could explore modifications to decrease the max drawdown of our method, resulting in more stable profitability. Additionally, since expectations are influenced by external factors such as financial reports or discussions on social media, future research could investigate incorporating this information into our model.
%
%
%
%
\bibliographystyle{splncs04}
\bibliography{mybibliography}

\end{document}